\documentclass[11pt,a4paper]{article}
\usepackage{arial}
\usepackage[T1]{fontenc}

\begin{document}

\title{ {\bf Understand assumptions and know uncertainties: Boscovich and the motion of the Earth}}

\author{Davor Krajnovi\'c\\
Institut f\"ur Astrophysik Potsdam (AIP)\\
An der Sternwarte 16, 14482 Potsdam\\
Email: dkrajnovic@aip.de\\}
\date{}
\maketitle

\begin{abstract}
\noindent The general prohibition of books advocating heliocentric theory put many progressive Jesuits in a difficult position. One of the most prominent Jesuit scientists of the 18th century, Rogerius Boscovich, was in particularly affected by conflicts between a beautifully simple theory of gravity by Newton, his Jesuit peripatetic education, Church doctrine and the lack of crucial experimental evidence for the motion of the Earth. I present the development of Boscovich's ideas prior to the lifting of the ban, and his retrospective considerations in later writings. These show that Boscovich's acceptance of the motion of the Earth was primarily driven by the existence of a working physical theory that also explained the motion of the Earth, and the lack of a consistent theory that supported any variation of a geocentric system.\looseness=-2
\end{abstract}

\section{A motivation for a change}

It is arguable how much of Rogerius Boscovich's\footnote{In English he is usually referred to as Roger Jospeh Boscovich, in Italian as Ruggiero Giuseppe Boscovich, while in his native Croatian, his name is spelled Ru\dj{}er Josip Bo\v skovi\'c (pronounced as ``Boshkovich''). Throughout this text, I will use the Latin version of his name, as this was how he signed most of his works.}  results and works influenced the coming generations of scientists and the overall knowledge, but his attitude to science and its methods are something that a present day scientist can still admire and take an inspiration from. He wrote once ``Illud homini est datum, ut minus erret, non ut omnino non erret''\footnote{Opera pertinentia ad opticam et astronomiam, vol. IV, Opusclum 4, Bassano 1785}, a simple formulation which all scientists should keep in mind at all times. The purpose of this paper is to outline Boscovich's struggle with a scientific framework of the stationary Earth in which he was brought up and which was commanded upon him by the authority of the Holy See and his own order, but which was not sustainable in the new age of rational enquiries and technological achievements. The path he took was not straightforward; sometimes he seemed to be lost, just to find a miraculous shortcut that can take him safely where he wanted to go. Later, when all obstacles were cleared from the highest instances, he acknowledged that miracles are not really needed in the world, just good understanding of natural philosophy. And it is this good understanding of the problems that should make Boscovich an unceasing inspiration for modern scientists. 

Boscovich's interest in the issue of the motion of the Earth was likely motivated by his reading of Newtown's Principa, and, possibly, more directly prompted by the most direct observation of the motion of the Earth, which happened just around the same time. At the end of 1725, James Bradley and Samuel Molyneux started observing $\gamma$ Draconis, a star that is seen near to the zenith at the latitude of London, and followed it over several months until they could show that the star did perform a peculiar annual motion. This was a result of a well set up experiment with the best possible technology of the day and a consequence that the measurements of positions of stars reached a precision of a few seconds of arc. A few years later, James Bradley published in the proceedings of the Royal Society (Bradley 1727) his explanation for this annual motion which he called the stellar aberration. These apparent motions of stars could only be explained as an effect of the motion of the Earth around the Sun, provided that the speed of light is finite (and constant).

This was a serendipitous discovery, but it was also a culmination of an arduous search for the proof that the Earth is in an annual motion around the Sun. At stake was the layout of the Universe, represented by two systems: the Copernican, or heliocentric, in which the Earth rotates around its axis and, together with other planets, around the Sun, and the system of Tycho Brache, or geo-heliocentric, in which the stationary Earth is in the centre of the Universe, around which the Sun rotates, but the planets rotate around the Sun, and hence around the Earth. By the mid 17 century the debate was summarised by the encyclopedic work of a Jesuit astronomer Giovanni Battista Riccioli. In the Book 9 of his ``Almagestum Novum'' (Riccioli 1651) he presented 126 arguments concerning the Copernican hypothesis of the motion of the Earth, 49 for and 77 against it. As pointed by Graney (2011), of those against, several deserve particular attention as they were related to questions in astronomy, physics and optics, three fields which interested Boscovich. They can be shortened and presented in the modern terminology in the following forms: i) if the Earth is in the annual motion around the Sun, then a star should be seen to have an apparent yearly motion and the angular extent of this motion, the parallax, should be measurable, unless the star is extremely distant; ii) if the Earth is in a diurnal motion, then it is a non-inertial frame with a uniform rotational motion and, as special forces occur in such frames, there should be observable effects, e.g. in the motion of cannon balls; iii) as stars seen through a telescope have a certain measurable sizes, but their parallaxes are below the observational measurements of the time, then their physical sizes (assuming the sizes seen in the telescope are real) should be extremely large (compared with the best estimates for the size of the Earth, Sun and the Solar system). These three major observational confirmations against the Copernican system were not disproven for a long time after Boscovich himself found a need to address the issue of Earth's motion\footnote{Bradley's discovery of the stellar aberration was not the discovery of the parallax. The first reliable measurements of the stellar parallax was done by Friedrich Wilhelm Bessel in 1838 (Bessel 1838). The main reason for this late discovery was in the necessary sub-arcsecond precision of the measurements (parallax of Bessels' star 61 Cygni is 0.28 seconds of arc).  The small parallax indeed puts the fixed stars at large distances, huge when compared with the distance Earth-Sun (for example the closest star, Proxima Centauri, is at a distance of 4.2 parsecs or 866312 astronomical units). A significant number of stellar parallaxes (>100 000) were only measured some 20 years ago with the European satellite mission Hipparcus (Perryman et al. 1997), while a new satellite mission of the European Space Agency, Gaia, scheduled to fly in 2013, should measure parallaxes of about 1 billion stars (Turon et al. 2005).

The physical understanding of the rotating frames and how to measure the effects were described in the 19. century by Gustave Coriolis (Coriolis 1835), while the most straightforward demonstration of the force and the diurnal rotation of Earth was presented in 1851 by Foucoult's Pendulum. 

Similarly, the understanding of the telescopic images of stars, or more precisely, the effects of the optical elements on the point like light sources, was not really understood until the work of George Biddel Airy and his description of the effect of the diffraction of light (Airy 1835), although astronomers such as Halley and Herschel seemed to understand that the sizes were not real (e.g. Graney \& Grayson 2011). 
}. 

Next to Bradley's strong observational evidence that the Earth is moving quite rapidly through space (approximately 30 km/s), there was also an intriguing discovery by Jean Richer in 1671 that the force of gravity is not the same at the Equator\footnote{More precisely, in Cayenne, at the latitude of 4$^o$ North.} and at moderate latitudes in Europe (e.g. Pannekoek 1961). Within the context of the Newtonian gravity, this implied a non-spherical shape of the Earth, which was flattened at the poles. A simple explanation of this phenomena could be achieved if the Earth is treated as a rotating spherical liquid and as MacLaurin, Huygens and Newton showed the daily rotation of the Earth could explain the measurements. Some years later, this result was challenged by the contradictory measurements of the Casini family who claimed the Earth is elongated at the poles. By the time of Boscovich, and after the return of the great French expeditions, one to Peru and the other to Lapland, the tide was turned again towards an Earth flattened at the poles (for a short history of meridian measurements see i.e. Heilbron 1993).

Clearly, the observational evidence for the motion of the Earth in the first half of the 18th century was not overwhelming, but there were significant indications. The strongest support, however, was not observational, but came from theory: Newton's gravitation. It had a great success explaining a number of natural phenomena (e.g. Kepler's laws and the shape of the Earth), but was incompatible with any proposed world system except the modified Copernican in which the paths of the planets are ellipses. Boscovich's conflict with the motion of the Earth should not be only viewed as a conflict between new and old ways within the Jesuit science, but also as an internal conflict of a scientist who has to find an agreement between a general lack of observational evidence for the motion of the Earth and an elegant theory which predicts it.

\section{Boscovich views on the rotation of Earth before 1757}

Boscovich, a natural philosopher and a Jesuit, was certainly in a complex situation. If by mid-18th century the debate between the cosmologies was moved somewhat away from the religious connotations, acknowledging the motion of Earth was still not officially allowed in the Catholic world. Natural philosophers, especially those that were Jesuits, had to be careful in expressing their thoughts on the subject. Boscovich's strategy of dealing with the situation evolved with time, and prior to the removal of the ban in 1757, it could be divided in two stages, with year 1747 as the significant transitional year in his considerations. 

\subsection{1738 - 1747}

In his early scientific writings, Boscovich seemed to be conservative regarding the motion of the Earth. Still, in this period Boscovich addressed the problem directly in five dissertations\footnote{As a comparison, Casanovas (1993) writes that in a period of similar length Boscovich's mentor Borgondio presented two dissertations on the problem of the motion of the Earth, one in 1714 (De telluris motu) and the other in 1725 (De situ Telluris).} which I will describe in more detail below. 

In the first work directly addressing the issue, ``De Aurora Boreali'' published in 1738, Boscovich modifies a theory by Jean-Jacques Dortous de Mairan (Marian 1731) abandoning the motion of the Earth as the cause for the polar light and constructs a theory in which the heating of the atmosphere is responsible for this extraordinary effect\footnote{Markovi\'c (1968, pp. 79-80)}. 

In the next year, Boscovich completed two didactic dissertations for public defences in the Collegium Romanum, which considered the shape of the Earth. He approached the question in two steps. In the first dissertation, ``De vererum argumentic pro telluris sphaericitate'', he put forward the traditional arguments that the Earth is spherical and pointed out that for a spherical shape of the Earth, it is necessary to assume that the Earth does not rotate. 

The second dissertation, ``Dissertatio de telluris figura'', addresses the problem more directly. Boscovich critically considered the then available measurements and the results on the shape of the Earth. Specifically, he considered three problems: i) what is necessary to measure a degree of a meridian, ii) how to determine the shape of the Earth if two measurements of a degree of a meridian exist and iii) the impact of several different theories of gravity (e.g. the force of gravity decreases with the square of the distance, constant gravity and gravity increasing linearly with the distance from the centre of the Earth). 

Boscovich's general conclusion is that the precision of the measurements is not sufficient to deduce the shape of the Earth, and, therefore its rotation. The main reasons are: the uncertainties in the derivation of the vertical and the direction of the force of gravity, due to the uneven structure and density of the surface of Earth, as well as the uncertainty in the determination of the zenith distance of the heavenly bodies due to unknown theory of refraction. As already pointed out by Markovi\'c (1968)\footnote{Markovi\'c (1968, p. 90)}, Boscovich gives an interesting explanation why the different length of the pendulum at the Equator and in Europe does not necessarily imply that the Earth is flattened by rotation. He suggests that ``the density of water is smaller than the density of the mountains, and that the density of bodies is larger in the colder regions than in warmer, so the gravity will be smaller on the island of Cayenne, which is in the middle of the ocean and in the tropics, than in the Mediterranean region, and even smaller than on the poles.''

Boscovich's interest in the theory of gravity continues in ``De inaequalitate gravitatis in diversis terrae locis'', published 1741, which repeats a number of issues mentioned in the works on the figure of the Earth from 1740. Boscovich again stressed the fact that the pendulum results could be explained with varying densities, and also restated the conclusion that not much can be inferred about the shape and rotation of the Earth. He justified this by certain (small) departures of the Newton's predictions from the observations of the pendulum motion. Furthermore, to prove his point on the uncertainties, although not really proposing a rival theory, he also considered a complex version of gravity which would allow a stationary Earth elongated at the poles\footnote{Markovi\'c (1968, p. 103)}.

In 1742, Boscovich wrote three dissertations related to astronomy. These dissertations outlined challenges to the accepted opinions about basic assumptions in astronomy such as: what is the nature of light, what is the nature of gravity and what is the theory of refraction. As in ``De inaequalitate gravitatis'', Boscovich criticised to the usage of unproven hypotheses in the scientific research in general. 

Of the three astronomical dissertations, the most interesting for the problem of the motion of the Earth is ``De annuis fixarum aberrationibus''\footnote{Other two are ``De observationibus astronomicis et quo pertingat earundem ceritudo'' and ``Disquisitio in universam astronomiam''.}, in which Boscovich confronted the observations of the (annual) motion of the Earth: the existence of various types of aberrations; the parallax of the fixed stars and Bradley's stellar aberration. 

Boscovich was very clear that he did not believe in the existing claims for the measurements of the parallax of the fixed stars. On the other hand, he acknowledged the existence of the aberration of the stellar light, and praised Bradley's explanation of this phenomena. Still, he put forward a question: is it really possible to explain the effect only by invoking the motion of the Earth, even if one assumes that the light is not spreading instantaneously? As one could expect within the context of the time and place, Boscovich's answer was negative. 

Here he did not stop only at the criticism of the existing theories but discussed a set of plausible theories which do and do not require the motion of the Earth. He addressed the following cases: there is no aberration but parallax exists, aberration exists, but there is no parallax, aberration and parallax exist, and there is no aberration, but the speed of light is infinite, or the light arrives to Earth in 3 months, or the light arrives to Earth in less than 3 months.

The first three cases assume the motion of the Earth, while in the other Boscovich considered the motionless Earth. Furthermore, in order to explain the aberration (and the possible parallax), he introduced a crucial modification to the system of Tycho: not only that the planets and comets follow the Sun motion around the Earth, but so do also the fixed stars. In the dissertation, Boscovich derived the apparent motions of the fixed stars in the given cases and gave practical advice on how to perform the observations. 

Boscovich concluded ``De annuis fixarum aberrationibus'' by stressing that ``results depend on hypotheses''\footnote{Markovi\'c (1968), p. 107}. The same thought features strongly in the third study of that year, ``Disquisitio in universam astronomiam''. Here  Boscovich expressed his views on what is a good and what is a bad hypothesis. A bad hypothesis is, essentially, one which can not be verified by the observations (as well as the trivial case, when there are clear counter arguments). In this dissertation, Boscovich, while claiming that Newton's theory of gravity is probably the best of all presented theories\footnote{Markovi\'c (1968), p. 122}, declares that it is still not a sufficiently good hypothesis, and brings forward two known critiques of the work: the fact that the fixed stars do not collapse on each other, and that it is not clear how particles act on each other over distance, or in other words, how gravity actually works. 

In this period Boscovich firstly ignores the motion of the Earth, then attempts an ad hoc improvement of Tycho's geo-heliocentric system, and finally, instead of addressing it more directly, starts systematically presenting Newtown's physics. Although the dissertations are, as required, negative towards it, and Boscovich seems to find a scientifically acceptable reasons to save the traditional Jesuit physics, he imposes on his students, year after year, a necessity to actually master Nawton's ideas, and consider the effects of the motion of Earth. Various authors (Markovi\'c 1968; Dadi\'c 1987; Casini 1993; Casanovas 1993) pointed out that Boscovich's motive for choosing these specific topics (the figure and the motion of the Earth, the theory of light and, more generally, the interplay of optics and astronomy) as subjects of the yearly dissertations which his students had to defend, was to introduce the Newtonian physics into the curriculum. As Casanovas (1993) points out, accepting the theory of gravitation of Newton, defines also the framework of the Universe. One should consider within this context also Boscovich's interests in measuring the length of the meridian or the first considerations of the pendulum clock in 1741 with brothers Jacuier and Le Seur in the gardens of their monastery Trinit\'a de'Monti, as well as his letter of 1 July 1741 to his brother Bo\v zo in Dubrovnik where he inquires if there is a good pendulum clock that shows seconds and whether Benedikt Stay\footnote{A compatriot who accomplished a significant career within the Roman Curia, later also collaborated with Boscovich in a unique popularisation of Newtonian (and Boscovich's) physics in ``Philosophiae recentioris vesibus traditae libir X" (Stay 1755).} could do the measurements\footnote{Markovi\'c (1968, p. 101)}. Boscovich is keenly interested to get his own data to test the Newtonian physics.

\subsection{1747 - 1757}

The second phase of Boscovich treatment of Earth's motion appears as a product of a great frustration. He was clearly enchanted by Newton's rigorous geometrical approach and the idea that a number of natural phenomena can be explained by a single physical law\footnote{Markovi\'c (1968, p. 121)}. This did not stop him criticising various aspects of the Newtonian physics, which, of course, will lead to his ``Philosophiae naturalis theoria'' (Boscovich 1758), but by this moment, he must have realised if Newtonian gravity was true, motions of the Earth, both annual and diurnal were inevitable. How to reconcile this progressive scientific thoughts with the doctrine of the Church and the guidelines of the Ratio Studiourum of the Collegium Romanum? His answer was to develop a scientifically driven hypothesis, acceptable to both the theologians and scientists\footnote{In ``Disquisitio in universam astronomiam'', Section 38; Markovi\'c (1968,  p .129)}. 

Boscovich's opinion that motion is relative, or that absolute motion is impossible to deduce or prove empirically, is what led him to the miraculous path out of the dilemma. Suppose there is a space which contains all visible Universe, the Sun, the Earth, the Moon, all the planets and all the fixed stars, whatever their distance from the Sun might be. If this stellar space, performs all the motions like the Earth, exactly but in an opposite direction, then the Earth will appear to stand still to an observer outside of this space\footnote{In ``De Cometis'' (Section 15,  p. 325), ``De aestu maris'' (Sections 69 and 71, pp. 33-34) and Markovi\'c (1968, p. 131).}. The omnipotent observer who can view the Universe from the absolute space\footnote{This motionless space is in principle equivalent to Newton's absolute space.}, will see Earth motionless, because the stellar space moves to cancel the Earth motions within it. Observers within this stellar space, however, will see the Earth moving, as they can not detect the motion of the space of which they are a part.

This was Boscovich's solution which he presented firstly in the dissertation in ``De cometis'', 1746, developed in more detail in ``De aestu maris'', 1747, and referenced in ``Dissertazionie della tenuit\'a della luce Solare'', 1747, ``Dissertationis de lumine'' 1748, ``Sopra il Turbine'' 1749, ``De literaria expeditione'', 1755, notes to Stay's ``Philosophiae recentioris'' 1755, and finally in ``De ineaqualitatibus'' 1756, the year before the ban on the Earth's motions was lifted. In the supplements to Noceti's poem ``De Iride et Aurora boreali''\footnote{``De Iride et Aurora boreali'', Note 81, pp. 116-117; Markovi\'c (1968, p. 129)}, Boscovich explained that he thought deeply about this problem and came to the conclusion, which might look arbitrarily constructed, but he was convinced it could not be proven wrong, in terms of logical arguments or experimental data. 

This metaphysical construction looked contrived even to Boscovich, but he forcefully supported it. He saw it as the only way that allowed him to use the motion of Earth in practice. He is explicit in this in ``De Cometis'', admitting that it is invented to keep the Earth stationary\footnote{``De Cometis'' (Section 19,  p. 327); Markovi\'c (1968, p. 134)}. Many years later, he goes as far as to say that it is infinitely improbable, but if the Creator wanted it, it would be possible\footnote{``Notice abr\'eg\'ee de l'Astronomie pour un Marin'',  Opera V, p 319; Markovi\'c (1968, p. 135)}. In a way, he is, as he himself says, after more than a century returning the whole discussion back to its theological origins. As Markovi\'c (1968) points out\footnote{Markovi\'c  (1968, p. 133)}, in ``De Cometis'' Boscovich explicitly mentions that the discussion which led to the prohibition of the motion of the Earth in 1615 was of theological nature. The question now is not whether Earth is moving or not, but whether the stellar space is the motionless absolute space or it moves with respect to the real absolute space. Crucially, for Boscovich the nature of the stellar space is the one that should be considered by the theologians, while the motion of the Earth can be accepted by the scientist. 

Boscovich's solution to the motion of the Earth problem seemed to remove all obstacles and allowed him to discuss any related topic. It allowed him to pursue Newtonian physics in all its aspects, but also to address topics in natural philosophy which he thought fundamental, such as the shape of the Earth through measurements of length of a meridian, and explaining the Jupiter-Saturn system by Newtonian gravity. While in the first period he might have looked reluctant to openly admit the new physics is a better physics, or perhaps even thought something could be done within the traditional framework, in the second period such considerations do not exist. Boscovich became a Newtonian, but one who is inspired by the new physics in ordered to take it further on. 

\subsection{Motion of the Earth in Boscovich's Literary writings}

Discussing the motion of the Earth was, however, not prohibited in a poetical context, or even in a strict mathematical context, as these both did not represent the reality, but a metaphorical and an ideal world, respectively. Boscovich used this opportunity and, as a member of the Accademia degli Arcadi since 1746, he promoted in multiple occasions the heliocentric system in his literary works. 

Not strictly literary notes to Noceti's ``De Iride et Aurora boreali'' were already mentioned as the vehicle for the popularisation of Boscovich's solution to the motions of the Earth. A year later, Boscovich used the northern lights again as an opportunity to express his opinion on the matter, now in a proper Arcadian prose description of the heliocentric system. In ``Dialoghi sull'aurora boreale", published in the edition of Giornale de' Letterati for year 1748, Boscovich clearly mentions that planets go around the Sun as the moons go around Jupiter, and that the attraction force works as ``lunghe redini incontrastabili''\footnote{Boscovich, Giornale de' Letterati, art. XXX, pp 298-299. Also in Casini (1993).}.  

The next literary publication of interest here was treated in detail by Martinovi\'c (1993), and concers an epigram of Boscovich published in the anthology of Roman arcadian poetry, ``Arcadium carmina pars altera'', in 1756. As Martinovi\'c points out, a year before the official removal of the ban by the Holy Office on the notion of the motion of the Earth, Boscovich declares that the  Earth is between Venus and Mars\footnote{The relevant part of the epigram  declares: ``Mercury closest is to Phoebus, Venus from the further foot/To Venus full close, the Earth's does steal a step./
While Earth with vast circuit Mars surrounds.'' (The full text of the Epigram in English translation of G. McMaster can be found in the Catalogue of the exhibition: ``Ru\dj{}er Bo\v skovi\'c again in his native Dubrovnik'', 15.05-15.09.2011, Martinovi\'c (2011).}. The emphasis on the metaphor of the Earth being between the Gods of Love and War, should certainly be taken into account, but more as an evidence of a poetically inspired of Boscovich. With this epigram Boscovich declares his heliocentric conviction again clearly and openly. In addition to the reference to the location of the Earth, in the following verse Boscovich explains the foundations of the heliocentric universe: ``Watch then with what order gold-tressed Sun / Sets ordered heavn'ly vault in circled course!''. It is the Sun that sets the heavens in order, presumably through the gravitational force of Sir Isaac Newton.  

It is possible that the recital of the epigram happened some time before, within the Accademia degli Arcadi. One could say that, as this was a poetical society, Boscovich support to the heliocentric universe, under a mask of the arcadian Numenio, were not considered as dangerous or grossly out of place, even if among its members there were cardinals and church officials. It can be imagined that by late 1740s the atmosphere in Rome had changed (as also alluded in his note to the Noceti's poem) sufficiently that he was able to start declaring his views more openly, or at least he was certain in a protective support from some part of the Church and Jesuit hierarchy to take his stance\footnote{In 1746 there are first official and negative grades given to Boscovich, which suggest a growing colision with the Jesuit hierarchy due to reach its peak with the case of Benvenuti in 1754 (e.g. Baldini 1993, Martinovi\'c 2011).}.
  
In this period one could also mention that Boscovich was intensively working on his major literary work: ``De Solis ac Luna Defectibus''. This didactic poem, a popular course on astronomy, was published for the first time in London in 1760 and dedicated to the Royal Society, but the first recitals of its stanzas happened much earlier. As pointed our by Martinovi\'c, Boscovich included the above mentioned epigram in the first book of the poem\footnote{``De Solis ac Lunae defectibus'', lib I, v510.}.

In this literature context, although as a side note, it is interesting to mention how was Boscovich perceived in the Italian (literary) circles. As Torbarina (1950) points out, Italian writer Giuseppe Baretti, in the 9th volume of ``La frusta letteraria''\footnote{G. Baretti: La frusta letteraria, published 1.02. 1764, a cura fi L. Piccioni, Laterza, Bari, 1912. In the same text Baretti wrote the memorable sentence: ``i quali sono tutti uomini ma non sono tutti Boscovich''.}, declares Boscovich as a famous (Papal) Astronomer who accepts the Copernican system; Boscovich was used as a representative of those progressive Catholic thinkers that supported the motion of the Earth.

\section{After the lift of the ban}

The clear statements of Numenio about the motion of the Earth and Boscovich's continuing wish to discuss the issue and a struggle to resolve it in a scientifically acceptable way, put little doubt on how welcomed must have been the printing of the new Index in 1757. Could there be any doubt in Boscovich's firm support of the heliocentric system prior to 1757, as some authors rose this concern (e.g. Dadi\'c 1987\footnote{Dadi\'c (1987, pp. 59-60)})? Could it be imagined that his Jesuit background and his religious beliefs influenced his understanding of the natural phenomena to the level that Boscovich never really wanted to accept the heliocentric system? The answer to these questions can be found in Boscovich's research and explicit statements on the topic after 1757. In particular, in his work on the telescope filled with water, his reprinting of ``De Cometis'' and further developments of the method of determination of comet trajectories and, finally, in a little textbook on astronomy ``Notice abr\'eg\'ee de l'Astronomie pour un Marin''. The printed versions of all these works appeared as part of his ``Opera pertinentia ad opticam et astronomiam'' in 1785, but they were results of earlier efforts.\footnote{Detailed discussion on the telescope filled with water and methods on determination of comet trajectories  were presented in Volume III, while the text book on astronomy in Volume V of the ``Opera''.}

\subsection{Terrestrial experiment with a telescope filled with water}

After the publication of the ``Philosophiae naturalis theoria'', Boscovich's main scientific interests seemed to have changed somewhat towards the optical phenomena, although he was clearly still very interested in the astronomical research. He was without doubt keenly interested in the nature of light, and the properties of light have a central role in the discussion on the motion of the Earth. For example, Bradley's explanation of the stellar aberration is only valid if the light propagates at a constant speed. Boscovich's proposal of an experiment with a telescope field with water had a purpose to distinguish between the existing theories of light, but his later addition of observing the light from a terrestrial source explicitly contained that fact that the Earth is moving. Boscovich believed that it is possible to measure the effect of aberration of terrestrial light source in a water telescope only if the telescope and the light source are oriented ``perpendicular to the radius of the ecliptic to the tangent of the Earth's annual motions.''\footnote{Opusculo III, Opera vol II, (1785), as translated into English in Proverbio (1993)}.

The importance of this point should not be overlooked. Boscovich's initial idea was to verify corpuscular nature of light and confirm Newton's theory of light, and to do so he was using as an assumption that the Earth is moving. In addition, he implicitly confirms that Bradley's stellar aberration is due to the annual motion of the Earth and finite speed of light. Boscovich's first idea of the telescope filled with water can be dated to 1766, while the experiment with a terrestrial light source was conceived while working on his collected works in Bassano in 1780s and at that time there was no need to spill any ink on explaining the obvious fact that the Earth is moving. 

\subsection{Comet trajectories}

The several treaties on the method of determination of comet trajectories are certainly in the centre of Boscovich's astronomical and mathematical research, but a note in the reprint of the original ``De Cometis'' dissertation in Opera is insightful. There Boscovich writes: ``Where the Earth's motion is mentioned in the discussion, the reader will have to take into consideration the time and place of the publication of the first edition. This was outlined during the annual discussion in Rome in 1746, before the clause forbidding all books which confirm the Earth's motion was cancelled and removed from the new printed index. This took place after thorough and thoughtful discussions.''\footnote{Text in English translation is taken from Fran\v si\'c (1987, p. 104).}  Furthermore Boscovich adds that the concept of the relative motions of the absolute and stellar space allowed him to keep the respect to his state as a Christian, as well as to separate the philosophical from the religious questions.\footnote{Opera, Vol III, pp 317-319, see also Casanovas (1993).}  

It is quite clear Boscovich wanted to distance himself from his contrived ideas since 1740. He also wanted to make sure the reader understood this was a construction which primarily enabled him to discuss the celestial mechanics within the new framework of the Newtonian physics, and gave him a great freedom and space to manoeuvre his scientific thoughts. If not for anything else, this theory should be considered remarkable for that alone.

\subsection{A short Astronomy textbook}

Boscovich's final statements on the world system are contained in ``Notice abr\`eg\'ee de l'Astronomie pour un Marin''\footnote{Franu\v si\'c (1987) quotes the following translations of the title into English: ``An Introduction to Astronomy for the Mariner'' (Markovi\'c 1959) and ``A short review of Astronomy for the mariner'' (Markovi\'c 1969). The text was also translated into German as ``Abriss der Astronomier mit R\"ucksicht auf ihre Verbindung mit der Schiffahrt'' (H.E. W. Eschenbach 1787).}, published as Opusculum 4 in the last volume of the Opera, but written in 1775 while he was still living in Paris. In the third chapter ``On Space Motion of Stars and Their Physical Causes'' Boscovich presents the Copernican and Tychonic systems. This is interesting as there is no reason why he should talk about the system of Tycho some 18 years after the removal of the heliocentric system from the Index. It almost implies he wanted to explain what is it that he meant about the issue. He might have been motivated to lessen the critique from some members of the Acad\'emie des sciences, but also by a sense for scientific rigour and the fact that he himself was not happy with his construction of the absolute and relative spaces in the 1740s. The note in ``De Cometis'' explains why it happened, and this section in the ``Notice'' presents Boscovich's explicit statement why one should really discard the system of Tycho and favour the heliocentric system. 

Boscovich starts by stating: ``Tycho's system, albeit much more complicated, explains all phenomena equally well as that of Copernicus, and all the reasons derived by Galileo from astronomical phenomena fail to prove the worth of Copernicus's system as against Tycho's''\footnote{Opera, Vol III, p. 301, English translation taken from Dadi\'c (1987).}. This is true and is in line with what Riccioli and Jesuits of 17th century concluded. Nonetheless, by 18th century the understanding of the world had improved and Boscovich was not satisfied. 

The crucial quote is the following: ``The arguments deriving from the successive propagation of light and from the physical causes of motion, available to us today and found to comply increasingly with the phenomena, necessarily imply the diurnal and annual motion of the Earth unless one accepts the assumptions which I developed thirty years ago and which provide the obvious although infinitely improbable possibility that the opposite might be true.''\footnote{Opera, Vol III, p. 301, English translation taken from Dadi\'c (1987, p. 60).}

This statement should be read within the context that the experimental evidence for the heliocentric systems were still not firmly established at that time. Certainly there was Bradley's stellar aberration, there were indications on a flattened shape of the Earth, but other confirmations, equally important and already pointed out more than a century before, were still missing. No parallax was measured and evidence of the existence of the rotating frame was not demonstrated yet. However, the theory of light, in particular, that it is propagating at a constant speed, and Newton's theory of gravity which explained the motion of heavenly bodies, are crucial aspects and the critical support for the heliocentric systems that existed by the end of the 18th century. 

Boscovich, as a good scientists, was skeptical of all explanations of phenomena that were not supported by experiments. Experiments, however, often depend on the technological achievements of the age. It took more than 200 years to find the first parallax, because it took so much time to achieve the necessary precision in instrument making. In the dissertations on astronomy from 1742, Boscovich outlines in detail all aspects of astronomy which are uncertain due to unsecured experimental results, or unverified assumptions. However, building a coherent theoretical system which explains the majority of other, unrelated, phenomena, can be sufficient for scientists to take it on as a working paradigm. A lack of such a theoretical systems to support contradicting phenomena is perhaps even stronger evidence for abandoning them. 

This is exactly the situation in which Boscovich saw the relative weights between heliocentric and geo-heliocentric systems. The system in which the Sun, the largest and the most massive body in the Solar system, is rotating around the Earth is not compatible with the most successful theory of gravity by Boscovich's times. The unequivocal support of the new physics, the heliocentric system and the motion of the Earth is furthermore supported by this sentence of an insightful physicist: ``In the celestial system there is nothing motionless unless there is [an] imagined point, called the common gravitational centre which all planets and comets, including the Sun, revolve around''\footnote{Opera, Vol III, pp. 304-305, English translation from Franu\v si\'c (1987).}

This is the reason why Boscovich from the beginning of 1740s repeatedly investigated Newton's gravity. He started conservatively with the first dissertations on the figure of Earth, but then he got his own hands dirty, figuratively and literary, trying to measure the deviations form the uniform gravitation with pendulum clocks, measuring the length of the meridian, pushing theoretically Newtonian physics to its limits and taking it beyond in multiple works until the publication of his ``Theoria''. Boscovich does not try to address Riccioli's arguments against the Copernican system; he is satisfied with having a testable hypothesis that mass exerts attractive force at large distances, which explains the working of the heavens. 

\section{Conclusions}

Boscovich's approach to the problem of the motion of the Earth was primarily scientific. It might have been motivated by his reading of Newton and the initial (negative) tone was almost certainly set by the formal requirement for its dismissal. It is plausible that his initial approach was to reconcile the old physics with the new discoveries (Markovi\'c 1968)\footnote{Markovi\'c (1968, p. 130)}, but quickly his attention was drawn fully to the new physics. His method of addressing the problem  was by presenting the new physics (and educating his students), and then looking for possible flaws and deficiencies. He did this in early dissertations by mid 1740s, at the very beginning of his research career. These were very fruitful early creations, as they motivated him to labour on various problems throughout his life and inspired his best works. 

There are many examples in contemporary science which could be compared with the battle between the world systems. In astrophysics, for example, it seems that 95\% of the matter in the Universe can not be seen directly and it is not made of the ordinary, baryonic matter (Ade et al. 2014). It is detected because the matter that is seen (e.g. stars or clouds of gas), in the outer regions of galaxies or in clusters of galaxies, dynamically behaves as if it is under the influence of gravitational motions of a large amount of invisible matter (e.g. van Albada et al. 1985). Astronomers call this unknown matter dark matter, and there is evidence that it indeed could exists (e.g. Trimble 1987), even though we do not understand what it is. 

There is another way of describing these observations. One could assume that the law of gravity changes in some particular manner, as to describe the dynamical properties of the visible matter, just as if there would be only visible matter (Milgrom 1983, Sanders \& McGaugh 2002). The choices are: to stick with the known theory of gravity and invent new type of matter, or to modify the law of gravity; after all Einstein (1915) did exactly that to Newton's theory. Which is the right choice? Perhaps it is still too early to say, even though 40 years of intensive research have already passed. Only by following Boscovich's bidding, if we understand the assumptions and know the uncertainties, can we hope to make advances in our knowledge of the natural world.

\vspace{+0.5cm}
\noindent {\bf References:}\\

\noindent Ade et al., 2014, {\it Planck 2013 results. XVI. Cosmological Parameters}, Astronomy \& Astrophysics, 571, 16

\noindent Airy, G. B., 1835, {\it On the Diffraction of an Object-glass with Circular Aperture}, Transactions of the Cambridge Philosophical Society, Vol. 5, p. 283-291.

\noindent Baldini, U, 1993, {\it Boscovich e la tradizione gesuitica in filosofia naturale: continuitˆ e cambiamento}, in in R.J. Boscovich, his life and scientific work, ed. P. Bursill-Hall, Istituto della enciclopedia Italiana, Rome.

\noindent Baretti, G, 1764, {\it La frusta letteraria}, published 1.02. 1764, a cura di L. Piccioni, Laterza, Bari, 1912.

\noindent Bessel, F. W,  1838, {\it On the parallax of 61 Cygni}, Monthly Notices of the Royal Astronomical Society, 4, 152, 

\noindent Boscovich, R, 1738, {\it De Aurora Boreali}, Ex typographia Antonii de Rubeis, Roma

\noindent Boscovich, R, 1739, {\it De vetrum argumentis pro telluris sphaericitate}, Typis Antonii de Rubeis, Roma

\noindent Boscovich, R, 1739, {\it Dissertatio de telluris figurea}, Typis Antoniii de Rubeis, Roma

\noindent Boscovich, R, 1740, {\it De motu corporum projectorum in spatio non resistente}, Typis Antonii de Rubeis, Roma

\noindent Boscovich, R, 1741, {\it De inaequalitate graitatis in diversis Terrae locis}, Typis Antonii de Rubeis, Roma

\noindent Boscovich, R, 1742, {\it Disquisitio in universam astronomiam}, Ex Typographia Komarek, Roma

\noindent Boscovich, R, 1742, {\it De observationibus astronomicis, et quo pertingat earundem certitudo}, Typis Antonii de Rubeis, Roma

\noindent Boscovich, R, 1742, {\it De annuis fixarum aberrationibus}, Ex typographia Komarek, Roma

\noindent Boscovich, R, 1743, {\it De motu corporis attracti in centrum immobile viribus decerescentibus in ratione distantiarum reciproca duplicata in spatiss non resistentibus},Typis Komarek, Roma

\noindent Boscovich, R, 1745, {\it De viribus vivis}, Sumptibus Venantii Monaldini Bibliopolae, Typis Komarek, Roma

\noindent Boscovich, R, 1746, {it De Cometis}, Ex Typographia Komarek, Roma

\noindent Boscovich, R, 1747, {\it Notae in auroram borealem. In Carolus Noceti, De Iride et Aurora Boreali Carmina}, Ex Typographia Palladis, excuddebant Nicolaus et marcus Palearini, Roma,  pp 89-127

\noindent Boscovich, R, 1747, {\it Dissertatio de maris aestu}, Ex Typographia Komarek, Roma

\noindent Boscovich, R, 1747, {\it Dissertazione della tenuitˆ della luce Solare}, Giornale de' Letterati, no. II, p.27, Roma

\noindent Boscovich, R, 1748, {\it Dissertationis de lumine pars prima}, Typis Antonii de Rubeis, Roma

\noindent Boscovich, R, 1748 {\it Dialoghi sull'aurora boreale}, in Giornale de' Letterati, Roma

\noindent Boscovich, R, 1756 {\it De inaequalitatibus quas Saturnus et Jupiter sibi mutuo videntur inducere praesertim circa tempus conjunctionis},  Ex Typographia Generosi Salomoni, Roma

\noindent Boscovich, R, 1758, {\it Philosophiae naturalis theoria redacta ad unicam legem virium in natura existentium}, In officina libraria Kaliwodiana, Vienna

\noindent Boscovich, R, 1765, {\it In planetarum dispositione Terra inter Martem et Veneram epigramma. In Arcadum carmina, pars altera},  Typographia Josephi \& Philippi de Rubeis, Roma, pp 214-215.

\noindent Boscovich, R, 1760, {\it De Solis ac Lunae defectibus}, Apud Andream Millar, in the Strand, et R. et J. Dodleios in Pall-mall, London, 1760 

\noindent Boscovich, R, 1785, {\it Opera perinentia ad opticam et astronomiam, I-V}, Venetiis apud Remondini, Bassano, 1785

\noindent Boscovich, R, 1785, {\it Notice abr\'eg\'ee de l'Astronomie pour un Marin, Opera perinentia ad opticam et astronomiam}, Vol. 5, Opusculum 4, Bassano 1785

\noindent Bradley, J, 1727, Phil. Trans. vol. 35, 637-661

\noindent Casanovas, J, 1993, {\it Boscovich's Early Astronomical Studies at the Collefio Romano}, in R.J. Boscovich, his life and scientific work, ed. P. Bursill-Hall, Istituto della enciclopedia Italiana, Rome.

\noindent Casini, P, 1993, {\it Boscovich and the Hypothesis Terra Motae}, in R.J. Boscovich, his life and scientific work, ed. P. Bursill-Hall, Istituto della enciclopedia Italiana, Rome.

\noindent Coriolis, G-G, 1835,  {\it Sur les \'equations du mouvement relatif des syst\`emes de corps},  J. De l'Ecole royale polytechnique 15: 144Ð154.

\noindent Dadi\'c, Z, 1987, {\it Ru\dj{}er Bo\v skovi\'c}, \v Skolska Knjiga, Zagreb

\noindent Doppelmayr, J, G.1742, {\it Atlas Coelestis}, Norimbergae, Heredi Homannianorum

\noindent Einstein, Albert, 1915, {\it Zur allgemeinen Relativit\"atstheorie}, Sitzungsberichte der K\"oniglich Preu$\beta$ischen Akademie der Wissenschaften, 778, 

\noindent Franu\v si\'c, B, 1991, {\it The analyss of Boscovich's astronomy for the mariner}, in Zbornik radova Me\dj{}unarodnog znastvenog skupa o Ru\dj{}eru Bo\v skovi\'cu, Dubrovnik, 5-7.10. 1987, Zagreb

\noindent Graney C. M 2012, {\it 126 Arguments Concerning the Motion of the Earth, as presented by Giovanni Battista Riccioli in his 1651 Almagestum Novum}, Journal for the History of Astronomy, volume 43 (2012), p.215-226

\noindent Graney, C. M. \& Grayson, Timothy P., 2011, {\it On the Telescopic Disks of Stars: A Review and Analysis of Stellar Observations from the Early Seventeenth through the Middle Nineteenth Centuries}, Annals of Science, Volume 68, 351-373.

\noindent Heilbron, J, 1993, {\it Meridiane and Meridians in Early Modern Science},  in R.J. Boscovich, his life and scientific work, ed. P. Bursill-Hall, Istituto della enciclopedia Italiana, Rome

%\noindent Komatsu,  2009, {\it Five-Year Wilkinson Microwave Anisotropy Probe Observations: Cosmological Interpretation}, The Astrophysical Journal Supplement, 180, 330, 

\noindent Maire, C \& Boscovich R, 1755, {\it De litteraria expeditione per ponitificam ditionem ad dimetiendos duos meridiani gradus et corrigendam mappam geographicam, jussu, et auspiciis Benedicti XiV. Pont. Max. suscepta a Patribus Societatis Jesu Christophoro maire et Rogerio Josepho Boscovich},  In Typographio Palladis, excudebant Nicolaus, et Marcus Palearini, Roma

\noindent Marian, J-J, D, 1731, {\it Traite physique et historique de l'Aurore Boreale}, Paris

\noindent Markovi\'c, \v Z, 1968, {\it Ru\dj{}e Bo\v skovi\'c I}, JAZU, Zagreb

\noindent Markovi\'c,\v  Z, 1969, {\it Ru\dj{}e Bo\v skovi\'c II}, JAZU, Zagreb

\noindent Martinovi\'c, I, 1993, {\it Epigrami Ru\dj{}era Bo\v skovi\'ca}, Dubrovnik 4, br. 3, pp. 93-120.

\noindent Martinovi\'c, I, 2011, {\it Ru\dj{}er Bo\v skovi\'c again in his native Dubrovnik}, exhibition 15.05-15.09.2011, Dubrova\v cki Muzeji, Dubrovnik

\noindent Milgrom, M, 1983, {\it A modification of the Newtonian dynamics as a possible alternative to the hidden mass hypothesis}, ApJ, 270, 365, 

\noindent Pannekoek, A, 1961, {\it A history of Astronomy}, Allen \& Unwin, London

\noindent Perryman, M.A.C  et al. 1997, {\it The HIPPARCOS Catalogue}, A\&A, 323, 49

\noindent Proverbio, E, 1993, {\it Boscovich's Project for Verifying Newton's Throey on the Nature of Light},  in R.J. Boscovich, his life and scientific work, ed. P. Bursill-Hall, Istituto della enciclopedia Italiana 1993, Rome.

\noindent Riccioli, G-B,1651, {\it Almagestum novum astronomiam veterem novamque complectens observationibus aliorum et propriis novisque theorematibus, problematibus ac tabulis promotam (Vol. IÐIII, 1651)}, Bononiae, Haeredis Victorij Benatij

\noindent Sanders, R \& McGaugh, S, 2002, {\it Modified Newtonian dyncamics as an alternative to dark matter}, Annual Review in Astronomy \& Astrophysics, 40, 263, 

\noindent Stay, B, 1755, {\it Philosopiae recentioris a Benedicto Stay versibus traditae libri X cum adnotationibus, et supplementis P.Rogerii Josephi Boscovich I-II}, Typis et sumptibus Nicolai, et Marci Palearini, Roma, I

\noindent Torbarina, J,1950, {\it Bo\v skvi\'c u krugu engleskih knji\v zevnika}, in Gra\dj{}a za \v zivot i rad Ru\dj{}era Bo\v skovi\'ca, 1, ed. \v Z Markovi\'c, Zagreb, JAZU, 1950

\noindent Trimble, V. 1987, {\it Existence and nature of dark matter in the universe}, Annual Review in Astronomy \& Astrophysics, 25, 425, 

\noindent Turon, C. 2005, {\it The Three-Dimensional Universe with Gaia}, ESASP, 576, ``Proceedings of the Gaia Symposium'',  (ESA SP-576)

\noindent van Albada et al, 1985, {\it Distribution of dark matter in the spiral galaxy NGC 3198}, Astrophysical Journal, 295, 305

\end{document}